\documentclass[prb,showpacs]{revtex4}% Physical Review B
\usepackage{graphicx}% Include figure files
\usepackage{dcolumn}% Align table columns on decimal point

\usepackage{bm}% bold math
\begin{document}

\title{The pulse and monochromatic light stimulation of semiconductor
quantum wells}
\author{I. G. Lang}
\affiliation{ A. F. Ioffe Physical-Technical Institute, Russian
Academy of Sciences, 194021 St. Petersburg, Russia}
\author{S. T. Pavlov}
\address{ P.N. Lebedev Physical Institute, Russian Academy of Sciences,
119991 Moscow, Russia; pavlov@sci.lebedev.ru}

\begin{abstract}
The light reflectance and absorbance are calculated for a quantum
well (QW) the width of which is comparable with the light wave
length. The difference of the refraction coefficients of the
quantum well and barriers is taken into account. The stimulating
pulse form is arbitrary. An existence of two closely situated
discrete excitation energy levels is supposed. Such energy level
pare may correspond to two magnetopolaron states in a quantizing
magnetic field perpendicular to the QW plane. The relationship of
the radiative and non-radiative damping is arbitrary. The final
results does not use the approximation of the weak Coulomb
interaction of electrons and holes.
\end{abstract}


\maketitle

\section{Introduction.}
 At the light monochromatic and pulse irradiation of a
quantum well there appear the characteristic peculiarities in the
 reflected and transmitted light waves carrying an information
about the energy spectrum and lifetimes of the electron
excitations [1-4] The most interesting results are being obtained
in the case of the discrete energy levels of the electronic
system. Such situation is being realized in a quantizing magnetic
field perpendicular to the QW plane or in the case of the
excitonic states.

We are interested in the case of two closely located energy levels
since at a pulse irradiation there appears an effect of the
oscillations of the intensities of the reflected and transmitted
light on the frequencies close to the difference of the energy
levels [5]. Such situation appears, in particular, in the case of
the magneto-phonon resonance [6], under condition
\begin{equation}
\label{l} \omega_{LO}=j\omega_{e(h)H},
\end{equation}
where $\omega_{LO} $ is the longitudinal optical (LO) phonon
frequency, $e$ is the electron charge, $\omega_{e(h)H} =
|e|H/cm_{e (h)} $ is the cyclotron frequency, $m _ {e (h)} $ is
the electron (hole) effective mass, $H$ is the magnetic field. The
number $j$ may be integer, what corresponds to a "classical"
magnetopolaron, or fractional number (weak magnetopolaron) [7, 8].

For the high quality quantum wells the radiative damping
$\gamma_r$ of the light absorption line may be comparable  with
the non-radiative damping $\gamma$ or exceed it. In such situation
it is necessary to take into account all the orders on the
electron - light interaction [5, 7, 9-26]. Our previous results
(besides [20, 21, 23 - 26]) are true for comparatively narrow QWs,
when it is satisfied the condition I
\begin{equation}\label{2}
    kd<<1,
\end{equation}
where $k$ is the module of the light wave vector, $d$ is the QW's
width. The calculations show that only the reflection and
absorption peak positions depend on the QW's width, but not their
height and shape. In all the results, corresponding to the
condition (2), the difference of the refraction indexes of the QW
and barrier's was not taken into account. We are going to show
that for the narrow QWs the refraction index
 $\nu$ of the QW's material is omitted from all the final results.
There remains only the refraction index  $\nu_1$ of the barrier's
material.

However, in the case of the wide enough QWs when
\begin{equation}\label{3}
    kd\geq 1,
\end{equation}
the difference of $\nu$  and  $\nu_1$ must be taken into account,
what follows from the results of [20, 21, 23 - 26]. Let us remind
that the QW's width $d$ is limited from above by the demand of the
size quantization of the electron movement along the axes $z$.

Let us estimate the role of the Coulomb interaction of electrons
and holes. A separation of the variables  ${\bf r}_\perp$ and $z$
in calculations of the wave functions is possible, if the Coulomb
interaction weakly influences the particle movement in $xy$ plane.
It happens if
\begin{equation}\label{4}
    a^2_{exc}>>a_H^2,
\end{equation}
where $a_{exc}=\hbar^2\epsilon_0/\mu e^2$ is the Wannier-Mott
exciton radius in the magnetic field absence, $\epsilon_0$ is the
static permeability, $\mu$ is the adjusted effective mass,
$a_H=c\hbar/|e|H$ is the magnetic length. Using the parameters of
GaAs from [27], we obtain (in \AA)
\begin{equation}\label{5}
   a_{exc}=146, \,\,a_H^{res}=57.2,
\end{equation}
where  $a_H^{res}$ corresponds to the resonant magnetic field
 $H_{res}=20.2 $T, satisfying the condition
 (1) of the magnetopolaron resonance for an electron at $j=1$. Accorging to (5),
 we obtain
$(a_H^{res}/a_{exc})^2\cong 0.154$, i.e., the condition (4) is
satisfied. The influence of the Coulomb interaction on the
movement of the electrons and holes in the $xy$ plane had been
considered in [28].

The Coulomb interaction influence on the particle movement along
the $z$ axes may be neglected, if
\begin{equation}\label{6}
    a_{exc}>d.
\end{equation}
Let us check the compatibility of the conditions  (6) and (3). For
GaAs, for example, the energy gap is $\hbar\omega_g\simeq 1.52
eV$, the stimulating light frequency must be a little more of that
value. The module $k=\omega_g\nu/c$ of the light wave vector
equals $k=2.60\times 10^5 cm^{-1}$. Using (5) and (6), we obtain
$$kd<0.38.$$ In order to extend our theory on the case of the more wide
QWs, when the Coulomb forces influence the movement of electrons
and holes along the $z$ axes, we do not concretize the exciton
envelope wave function  $\phi(z)$ until obtaining the final
results.

\section{The stimulating field.}

The permeability inside of the QW is $\varepsilon=\nu^2$ , in
barriers $\varepsilon_1=\nu_1^2$ (Fig.1), where  $\nu, \nu_1$ are
the refraction coefficients. The electromagnetic wave propagates
along the $z$ axes perpendicular to the QW's surface (the $xy$
plane). Its electric field is as follows
\begin{equation}\label{7}
    \mathbf{E}_{0}(\mathbf{r},t) = E_{0}\mathbf{e}_{l}
\int_{-\infty}^{\infty}d\omega\,e^{-i\omega(t-z\nu_1/c)}\mathcal{D}_{0}(\omega)
+ c.c.,
\end{equation}
where $E_0$ is the scalar amplitude, $\mathbf{e}_{l}$ is the
polarization vector, $c$ is the light speed in vacuum. The
function $D_0(\omega)$ determines the shape of the stimulating
pulse and may be chosen as [5, 17-19,21]
$$\mathcal{D}_0(\omega)={i\over 2\pi}\left[{1\over
\omega-\omega_l+i\gamma_{l1}/2}-{1\over
\omega-\omega_l-i\gamma_{l2}/2}\right].$$ Under condition
$\gamma_{l1}=\gamma_{l2}=\gamma_{l}$ the light pulse is
symmetrical [5,18,19,21], at $\gamma_{l2}\rightarrow\infty$ it is
asymmetrical and has a very steep front [16,17,19]. At
$\gamma_l\rightarrow 0$ we obtain
$$\mathcal{D}_0(\omega)=\delta(
\omega-\omega_l),$$ what corresponds to the monochromatic
irradiation.

 \section{The electronic excitations in a QW.}
A QW is in a zero or quantizing magnetic field perpendicular to
the QW's surface at a zero temperature. The quasi-momentum matrix
elements ${\bf p}_{cv}$, characterizing the inter-band transitions
of an electron, are essential in the theory. As earlier [5,
7,16-22] we use the following model. The vectors ${\bf p}_{cv}$
for two sorts of excitons (with indexes I and II) are as it
follows
\begin{equation}\label{8}
{\bf p}_{cvI}=p_{cv}({\bf e}_{x}-i{\bf e}_{y})/\sqrt{2}, ~~~{\bf
p}_{cvII}=p_{cv}({\bf e}_{x}+i{\bf e}_{y})/\sqrt{2},
\end{equation}
where ${\bf e}_{x}$  and  ${\bf e}_{y}$ are the unite vectors
along the axis $x$ and $y$, respectively, $ p_{cv}$ is a real
value. The model corresponds to the excitons with a participation
of the heavy holes in the semiconductors with the zinc blend
structure if the axes  $z$ is directed along the 4-th order axes
[29,30]. Our results are applicable also to the excitons
$\Gamma_6\times \Gamma_7$ with a participation of the holes from
the valence band splitted by the spin-orbital interaction at an
arbitrary direction of the $z$ axes relatively the
crystallographic axis [31].

If the circular polarization vectors of the stimulating light
$${\bf e}_{l}=({\bf e}_{x}\pm i{\bf e}_{y})/\sqrt{2},$$
 are used, the conservation property of the polarization vector is performed:
$$\sum_{I,II}{\bf p}_{cv}^*({\bf e}_l{\bf p}_{cv})
=\sum_{I,II}{\bf p}_{cv}({\bf e}_l{\bf p}_{cv}^*)={\bf
e}_lp_{cv}^2.$$

The form of the wave function $F_\rho({\bf r})$ of the electronic
excitation at  ${\bf r}_e={\bf r}_h={\bf r}$ in the effective mass
approximation is essential in the theory.  ${\bf r}_e({\bf r}_h)$
is the electron (hole) radius vector [22]. If an excitation is
formed by the pair magnetopolaron - hole, the wave functions
contain the phonon components [8,32]. In such a case the value
$F_\rho({\bf r})$ is determined as follows: we multiply the
function from the left at a phonon vacuum  $\langle 0|$ and assume
${\bf r}_e={\bf r}_h={\bf r}$ .Such a method is justified by the
fact that at an exciton creation by light or at a light
annihilation of an exciton with a participation of a
magnetopolaron, it is essential only the term in the polaron wave
function which does not contain the phonons.

Let us assume that the variables  ${\bf r}_{\perp}$  and $z$ may
be separated, i.e.,
\begin{equation}\label{9}
    F_\rho({\bf r})=Q_\pi({\bf r}_{\perp})\phi_\chi(z).
\end{equation}
The separation of variables is possible, if the Coulomb
interaction of electrons and holes influences weakly at their
motion in the $xy$ plain.

\section{The induced current density.}

In order to calculate the electric fields on the left and on the
right of the QW and the field inside of the QW, it is necessary,
first of all, to calculate the density of the electric current
induced inside of the QW. If we use (8), we obtain the average
current density inside of the QW [22]
\begin{equation}\label{10}
    j_{1\alpha}={ie\over8\pi^2}\sum_\rho\gamma^0_{r\pi}\phi_\chi(z)\int_{-\infty}^\infty
    d\omega e^{-i\omega
    t}\int_{-d/2}^{d/2}dz^\prime\phi_\chi(z^\prime)E_\alpha(z^\prime,
    \omega)\left({1\over\omega-\omega_\rho+i\gamma_\rho/2}+
    {1\over\omega+\omega_\rho+i\gamma_\rho/2}\right),
\end{equation}
where $\rho$  is the index of the electronic excitation in the QW
 (combining the indexes
$\pi$ and $\chi$), $\hbar\omega_\rho$   is the excitation energy
counted from the ground state energy, $\gamma_\rho$ is the
non-radiative damping.

The value  $\gamma^0_{r\pi}$   is a factor in the radiative
damping expression. The radiative damping is determined as [22,23]
\begin{equation}\label{11}
    {\tilde
    \gamma}_{r\rho}=\gamma_{r\pi}|R_\chi(\omega_\rho\nu/c)|^2,
\end{equation}
where
\begin{equation}\label{12}
    \gamma_{r\pi}=\gamma_{r\pi}^0/\nu,
\end{equation}
\begin{equation}\label{13}
    R_\chi(k)=\int_{-d/2}^{d/2}dz\phi_\chi(z)e^{-ikz}.
\end{equation}
%.
The values  $\gamma_{r\pi}$  for an exciton consisting of a
magnetopolaron and hole were calculated in [7]. For some another
excitations they are given in [22].

Equation (10) may be obtained from  (46) from [22], if one takes
into account (12) and restricts the integration on $z^\prime$ by
the limits $-d/2$   and $d/2$  , i.e., neglects by the tunnel
penetration of the electronic excitations into the barrier
 (what, strictly speaking, corresponds to the infinitely deep QW).

\section{ The equation for the electric field
inside of the QW.}

Let us use the relationship
$${\bf E}(z, t)=-{1\over c}{\partial{\bf A}(z, t)\over\partial t},~~~\varphi=0,$$
where ${\bf A}(z, t)$   and  $\varphi$   are the vector and scalar
potentials, respectively. In the case of the model (8) the current
(10) is transverse one.- The $z$ -component is absent, the induced
charge density equals zero, hence it follows the condition
$\varphi=0$.

We start from the equation for the vector potential (see, for
instance, [33, p. 439])
\begin{equation}\label{14}
{\partial^2{\bf A}\over\partial z^2}-{\varepsilon\over
c^2}{\partial^2{\bf A}\over\partial t^2}={4\pi\over c}{\bf j}(z,
t)
\end{equation}
Let us wright the electric field as it follows {\footnote{The
partition of the RHS of (15) at the main and conjugated
contributions we proceed so in order to obtain equation (26) for
the exciting field.} }
\begin{equation}\label{15}
    {\bf E}(z, t)={{\bf e}_l\over2\pi}\int_{-\infty}^\infty
    d\omega E^{-i\omega t}\mathcal{E}(z, \omega)+c.c..
\end{equation}
Then with the help (10), (14) and (15) we obtain
\begin{equation}\label{16}
    {\partial^2\mathcal{E}(z, \omega)\over\partial z^2}+{\omega^2\nu^2\over
    c^2}\mathcal{E}(z, \omega)={\omega\over
    c}\sum_\rho\gamma_{r\pi}^0\phi_\chi(z)\int_{-d/2}^{d/2}dz^\prime\Phi_\chi(z^\prime)\mathcal{E}(z^\prime,
    \omega)\left({1\over\omega-\omega_\rho+i\gamma_\rho/2}+
    {1\over\omega+\omega_\rho+i\gamma_\rho/2}\right).
\end{equation}

\section{The case of two closely located energy levels.  The electric field
inside of the QW.}

Let us limit the sum in (16) by two terms with numbers: $i=1$ and
$i=2$. It is possible if the energy levels 1 and 2 are closely
located and others levels are situated at $\Delta\omega$ far away,
thus
$$\gamma_{1(2)}<<|\Delta\omega|,~~~\gamma_{r1(2)}<<|\Delta\omega|.$$
Let us consider a case when the functions  $\phi_i(z)$ for two
excitations coincide, i.e.,
\begin{equation}\label{17}
    \phi_1(z)=\phi_2(z)=\phi(z).
\end{equation}
For instance, such situation is realized for two closely located
energy levels of an exciton in the magnetopolaron effect [12-14].
Under condition (17) equation (16) results in
\begin{equation}\label{18}
    {\partial^2\mathcal{E}(z, \omega)\over\partial z^2}+k^2\mathcal{E}(z,
    \omega)=2k_0\phi(z)M(\omega)\mathcal{B}(\omega),
\end{equation}
where the designations
\begin{eqnarray}\label{19}
k={\omega\nu\over c},~~~k_0={\omega\over c},~~~M(\omega)
=\int_{-d/2}^{d/2}dz\phi_\chi(z)\mathcal{E}(z,
    \omega),\nonumber\\
\mathcal{B}(\omega)=\sum_{i=1,2}{\gamma_{ri}^0\over 2}
\left({1\over\omega-\omega_i+i\gamma_i/2}+
    {1\over\omega+\omega_i+i\gamma_i/2}\right)
\end{eqnarray}
are introduced. According to [34, p. 85], the solution of the
equation
$${\partial^2y\over\partial z^2}+k^2y=f(z)$$ is as
follows
\begin{equation}\label{20}
  y=C_1\cos kz+C_2\sin kz +{1\over k}\int_{z_0}^{z}dz^\prime f(z^\prime)\sin
k(z-z^\prime).
\end{equation}
Using (20), we obtain the Fourier transform of the electric field
inside of the QW
\begin{equation}\label{21}
    \mathcal{E}(z,
    \omega)=Ae^{ikz}+Be^{-ikz}-{i\over\nu}\mathcal{F}_k(z)M(\omega)\mathcal{B}(\omega),
\end{equation}
where the designation
\begin{equation}\label{22}
    \mathcal{F}_k(z)=e^{ikz}\int_{-d/2}^{z}dz^\prime\phi(z^\prime)e^{-ikz^\prime}+
e^{-ikz}\int_{z}^{d/2}dz^\prime\phi(z^\prime)e^{ikz^\prime}
\end{equation}
is introduced. Let us multiply (21) at  $\Phi(z)$ and integrate
from   $-d/2$ to $d/2$. We obtain the equation for the value
$M(\omega)$, the solution of which is as follows
\begin{equation}\label{23}
    M(\omega)={AR^*(k)+BR(k)\over
    1+(i/\nu)\mathcal{B}(\omega)J(k)},
\end{equation}
where
$$J(k)=\int_{-d/2}^{d/2}dz\Phi(z)\mathcal{F}_k(z),$$
$R(k)$ is determined in (13). Finally, having substituted (23)
into (21), we obtain that inside of the QW
\begin{equation}\label{24}
    \mathcal{E}(z, \omega)=Ae^{ikz}+Be^{-ikz}-{i\over\nu}\mathcal{F}_k(z)\mathcal{B}(\omega)
    {AR^*(k)+BR(k)\over
    1+(i/\nu)\mathcal{B}(\omega)J(k)},
\end{equation}
where $A$  and $B$ are the constants which must be determined,
using the boundary conditions on the edges of the QW.

\section{ The electric field on the left, on the right
and inside of the QW.}

It is obviously that the Fourier components
\begin{equation}\label{25}
\mathcal{E}_{left}(z, \omega)=\mathcal{E}_0(z,
    \omega)+\Delta\mathcal{E}_{left}(z,
    \omega),~~~\Delta\mathcal{E}_{left}(z,
    \omega)=Le^{-ik_1z},~~~\mathcal{E}_{right}(z, \omega)=Re^{ik_1z}
\end{equation},
correspond to the electric field outside of the QW. Here
$k_1=\omega\nu_1/c$, $L$ and $R$ are the constants,
\begin{equation}\label{26}
    \mathcal{E}_0(z,
    \omega)=2\pi E_0\mathcal{D}_0(\omega)e^{ik_1z}.
\end{equation}
At  $z=-d/2$  and  $z=d/2$  the magnetic field and the tangential
component of the electric field must be continuous. Since in the
case of the model (8) the longitudinal components of the fields
(along the $z$ axes) are absent, the boundary conditions may be
written as four equations
\begin{eqnarray}\label{27}
\mathcal{E}_{left}(-d/2, \omega)&=&\mathcal{E}(-d/2, \omega),\nonumber\\
\mathcal{E}_{right}(d/2, \omega)&=&\mathcal{E}(d/2, \omega),\nonumber\\
{d\mathcal{E}_{left}(z, \omega)\over
dz}\Big|_{z=-d/2}&=&{d\mathcal{E}(z, \omega)\over dz}\Big|_{z=-d/2},\nonumber\\
{d\mathcal{E}_{right}(z, \omega)\over
dz}\Big|_{z=d/2}&=&{d\mathcal{E}(z, \omega)\over dz}\Big|_{z=d/2}.
\end{eqnarray}
Having substituted (24) and (25) in (27),we solve the system of
four equation relatively of the constants A, B, L и R. As a result
we obtain the final expressions for the Fourier transforms of the
electric fields
\begin{eqnarray}\label{28}
    \mathcal{E}_{left}(z, \omega)=2\pi E_0\mathcal{D}_0(\omega)e^{ik_1z}
    +2\pi
    E_0\mathcal{D}_0e^{-ik_1z}Z^{-1}e^{-ik_1d}\Big\{(1-e^{-i2kd})(\zeta^2-1)\nonumber\\
    -ig(\omega)\Big[e^{-ikd}\Big((\zeta-1)^2R^2(k)+(\zeta+1)^2R^{*2}(k)\Big)
    +2(\zeta^2-1)|R(k)|^2\Big]\Big\},
\end{eqnarray}
\begin{equation}\label{29}
\mathcal{E}_{right}(z, \omega)=2\pi
E_0\mathcal{D}_0(\omega)e^{ik_1z}
    Z^{-1}e^{-i(k+k_1)d}\zeta\Big[1-ig(\omega)|R(k)|^2\Big],
\end{equation}
\begin{eqnarray}\label{30}
    \mathcal{E}(z, \omega)=4\pi E_0\mathcal{D}_0(\omega)Z^{-1}e^{-i(k+k_1)d/2}
    \Big\{e^{ikz}\Big[e^{-ikd}(\zeta+1)
    +i g(\omega)(\zeta-1)R^2(k)\nonumber\\
    +e^{-ikz}(\zeta-1)[1-i g(\omega)|R(k)|^2]
-i\mathcal{F}_k(z)g(\omega)[e^{-ikd}(\zeta+1)R^*(k)+(\zeta-1)R(k)]\Big]\Big\}.
\end{eqnarray}
We introduced the following designations
\begin{equation}\label{31}
    \zeta=\nu/\nu_1,~~~g(\omega)=\sum_{i=1,2}{(\gamma_{ri}^0/2\nu)L_i(\omega)\over
    1+iJ(k)\sum_{i=1,2}(\gamma_{ri}^0/2\nu)L_i(\omega)},
\end{equation}
\begin{equation}\label{32}
    L_i(\omega)={1\over\omega-\omega_i+i\gamma_i/2}+
    {1\over\omega+\omega_i+i\gamma_i/2},
\end{equation}
\begin{equation}\label{33}
    Z=e^{-2ikd}(\zeta+1)^2-(\zeta-1)^2+ig(\omega)\Big\{e^{-ikd}(\zeta^2-1)[R^2(k)+R^{*2}(k)]
    +2(\zeta-1)^2|R(k)|^2\Big\}.
\end{equation}

\section{The limit of the absence of electronic excitations.}

Assuming
$$\gamma_{r1}^0=\gamma_{r2}^0=0,$$
we move up to the situation, when light normally incident on a
surface of a plane-parallel plate is reflected due to the
difference of the refraction coefficients of the plate and medium.
Let us stress that in such a case there is no any limitations on
the plate depth $d$ relatively of the light wave length. With the
help of (28) - (30) we obtain the following results for the
Fourier transforms of the electric fields (compare with formulas
from [35, p. 412])
\begin{equation}\label{34}
    \mathcal{E}_{left}^0(z, \omega)=2\pi E_0\mathcal{D}_0(\omega)e^{ik_1z}
+2\pi
E_0\mathcal{D}_0(\omega)e^{-ik_1z}Z_0^{-1}e^{ik_1d}(1-e^{-i2kd})(\zeta^2-1),
\end{equation}
\begin{equation}\label{35}
    \mathcal{E}_{right}^0(z, \omega)=8\pi E_0\mathcal{D}_0(\omega)e^{ik_1z}
Z_0^{-1}\zeta e^{-i(k+k_1)d},
\end{equation}
\begin{equation}\label{36}
    \mathcal{E}^0(z, \omega)=4\pi E_0\mathcal{D}_0(\omega)e^{-i(k+k_1)d/2}
Z_0^{-1}\Big[ e^{ikz}e^{-ikd}(\zeta+1)+e^{-ikz}(\zeta-1)\Big],
\end{equation}
where
$$Z_0=e^{-2ikd}(\zeta+1)^2-(\zeta-1)^2.$$
Let us note that equations (34) - (36) may be used for an
investigation of the transmission and reflection of light pulses
from the plane of the transparent dielectric plates, if the
function
 $\mathcal{D}_0(\omega)$ describes the stimulating pulse continuation and shape.

\section{The limit of equal refraction indexes of the barriers and QW.}

In the case of $\nu=\nu_1$ from (28) - (30) we obtain the results
\begin{equation}\label{37}
    \mathcal{E}_{left}^{\nu=\nu_1}(z, \omega)=2\pi E_0\mathcal{D}_0(\omega)e^{ikz}
-2\pi E_0\mathcal{D}_0(\omega)e^{-ikz}g(\omega)R^{*2}(k),
\end{equation}
\begin{equation}\label{38}
    \mathcal{E}_{right}^{\nu=\nu_1}(z, \omega)=2\pi E_0\mathcal{D}_0(\omega)e^{ikz}
-2\pi E_0\mathcal{D}_0(\omega)e^{ikz}g(\omega)|R(k)|^2,
\end{equation}
\begin{equation}\label{39}
    \mathcal{E}^{\nu=\nu_1}(z, \omega)=2\pi E_0\mathcal{D}_0(\omega)\Big[e^{ikz}
-i\mathcal{F}_k(z)g(\omega)R^{*}(k)\Big],
\end{equation}
Using the definition (31) of the function  $g(\omega)$, as well as
the relationship
$$J(k)=|R(k)|^2+iq(k),~~~~q(k=0)=0,$$
we obtain from (37) and (38)
\begin{equation}\label{40}
    \mathcal{E}_{left(right)}(z, \omega)=\mathcal{E}_0(z, \omega)+\Delta\mathcal{E}_{left(right)}(z,
    \omega),
\end{equation}
\begin{equation}\label{41}
   \Delta \mathcal{E}_{left}^{\nu=\nu_1}(z, \omega)=2\pi
   E_0\tilde{\mathcal{D}}_\nu(\omega)e^{-i(kz-\alpha)},
\end{equation}
\begin{equation}\label{42}
   \Delta \mathcal{E}_{right}^{\nu=\nu_1}(z, \omega)=2\pi
   E_0\tilde{\mathcal{D}}_\nu(\omega)e^{ikz},
\end{equation}
where
\begin{eqnarray}\label{43}
    e^{i\alpha}={R^{*}(k)\over R(k)},\nonumber\\
\tilde{\mathcal{D}}_\nu(\omega)=-i\mathcal{D}_0(\omega){(\tilde{\gamma}_{r1}/2)L_1(\omega)
+(\tilde{\gamma}_{r2}/2)L_2(\omega)\over
1+i[(\tilde{\gamma}_{r1}/2)L_1(\omega)
+(\tilde{\gamma}_{r2}/2)L_2(\omega)]-\Delta_1L_1(\omega)-\Delta_2L_2(\omega)},
\end{eqnarray}
\begin{equation}\label{44}
    \tilde{\gamma}_{ri}={\gamma_{ri}^0\over\nu}|R(k)|^2,
\end{equation}
what coincide with the definition (11) in the approximation
$\omega=\omega_i$,
\begin{equation}\label{45}
    \Delta_i={\gamma_{ri}^0\over 2\nu}q(k).
\end{equation}
Let us note that in equation (32) for $L_i(\omega)$ one has to
neglect the second term (in order to avoid an exceeding
precision), what justifies the approximation
$\omega\simeq\omega_i$.

In order to make clear the physical sense of the values
$\tilde{\gamma}_{ri}$ and $\Delta_i$, let us go to the case of the
single energy level supposing  $\gamma_{r2}^0=0$.  Neglecting by
the non-radiative term from $L_1(\omega)$, we obtain

\begin{equation}\label{46}
    \tilde{\mathcal{D}}_{\nu 1}(\omega)=-i\mathcal{D}_0(\omega){\tilde{\gamma}_{r1}/2\over
    \omega-(\omega_1+\Delta_1)+i(\gamma_1+\tilde{\gamma}_{r1})/2},
\end{equation}
whence it follows that $\tilde{\gamma}_{r1}$  is the radiative
lifetime, $\Delta_1$ is the correction to the excitation energy.
However, if two energy levels are located closely enough, they
influence each other. Equation (43) may be transformed into
\begin{equation}\label{47}
\tilde{\mathcal{D}}_{\nu}(\omega)=-i\mathcal{D}_0(\omega)\left({\hat{\gamma}_{r1}/2\over\omega-\Omega_1
+iG_1/2}+{\hat{\gamma}_{r2}/2\over\omega-\Omega_2 +iG_2/2}\right),
\end{equation}
where $\hat{\gamma}_{ri}$, $\Omega_i$ and $G_i$  are the
"renormalized" values.

In the case of QWs and under condition $kd<<1$ we have
\begin{eqnarray}\label{48}
    R(k)\simeq\int_{-d/2}^{d/2}dz \phi(z)=C,~~~\mathcal{F}_k(z)\simeq
    C,~~~J(k)\simeq C^2,\nonumber\\
    q(k)\simeq 0,~~~\Delta_1\simeq\Delta_2\simeq 0,~~~
    \tilde{\gamma}_{ri}\simeq{\gamma_{ri}^0\over\nu}C^2,
    ~~~e^{i\alpha}=1.
\end{eqnarray}
Under condition $\nu=\nu_1$ we obtain from (43) for narrow QWs
\begin{equation}\label{49}
\tilde{\mathcal{D}}_{n\nu}(\omega)=-i\mathcal{D}_0(\omega){C^2\sum_{i=1,2}
(\gamma_{ri}^0/2\nu)L_i\over
1+iC^2\sum_{i=1,2}(\gamma_{ri}^0/2\nu)L_i},
\end{equation}
where the index $n$ corresponds to a narrow QW.

\section{The narrow QW. The different refraction coefficients in the
barriers and QW.}

Let us apply (48) and assume
$$e^{-ikd}\simeq e^{-ik_1d}\simeq 1.$$
Then we obtain from (28) - (29)
\begin{equation}\label{50}
     \mathcal{E}_{n~left}(z, \omega)=2\pi
     E_0\mathcal{D}_0(\omega)e^{ik_1z}-2\pi iE_0\mathcal{D}_0(\omega)e^{-ik_1z}
     {g_n(\omega)C^2\over
     1+ig_n(\omega)C^2(\zeta-1)},
\end{equation}
\begin{equation}\label{51}
     \mathcal{E}_{n~right}(z, \omega)=2\pi
     E_0\mathcal{D}_0(\omega){1-ig_n(\omega)C^2\zeta\over
     1+ig_n(\omega)C^2(\zeta-1)}=const,
\end{equation}
\begin{equation}\label{52}
     \mathcal{E}_{n}(z, \omega)=2\pi
     E_0\mathcal{D}_0(\omega)e^{ik_1z}{1-ig_n(\omega)C^2\over
     1+ig_n(\omega)C^2(\zeta-1)},
\end{equation}
where
\begin{equation}\label{553}
g_n(\omega)= \sum_{i=1,2}{(\gamma_{ri}^0/2\nu)L_i
    \over 1+iC^2\sum_{i=1,2}(\gamma_{ri}^0/2\nu)L_i}
\end{equation}
Having substituted (53) into (50) -(52) and substituted the
elementary transformations, we obtain the fields on the left and
on the right of the QW (40), where
\begin{equation}\label{54}
    \Delta\mathcal{E}_{left~ n}(z, \omega)=2\pi
     E_0\tilde{\mathcal{D}}_{n\nu_1}(\omega)e^{-ik_1z},
\end{equation}
\begin{equation}\label{55}
    \Delta\mathcal{E}_{right~ n}(z, \omega)=2\pi
     E_0\tilde{\mathcal{D}}_{n\nu_1}(\omega)e^{ik_1z},
\end{equation}
and the field inside of the QW
\begin{equation}\label{56}
    \mathcal{E}_{n}(z, \omega)=\mathcal{E}_{0}(z=0, \omega)+2\pi
     E_0\tilde{\mathcal{D}}_{n\nu_1}(\omega).
\end{equation}
where
\begin{equation}\label{57}
    \tilde{\mathcal{D}}_{n\nu_1}(\omega)=-i\mathcal{D}_0(\omega)C^2\sum_{i=1,2}{(\gamma_{ri}^0/2\nu_1)L_i
    \over 1+iC^2\sum_{i=1,2}(\gamma_{ri}^0/2\nu_1)L_i},
\end{equation}
what distinguishes from the result (49) for narrow QWs at
$\nu=\nu_1$ by the substitution of $\nu$ by $\nu_1$. It means that
at $\nu\neq\nu_1$ for narrow QWs ($kd<<1$) the radiative damping
contains the barrier refraction coefficient $\nu_1$, and the QW
refraction coefficient  $\nu$ does not appear at all! The physical
sense of the result is quite clear: Under condition $kd<<1$ we can
go to the limit $d\rightarrow 0$, when the QW's material is
absent, but there exists the induced current corresponding to the
exciton creation transitions.

Thus, it is proved  that our results for the narrow QWs for the
monochromatic and pulse irradiation [5,7,16-19] are true not only
at  $\nu=\nu_1$ , but also at $\nu\neq\nu_1$, since our formulas
contain only the barrier refraction coefficienyt. \footnote{In
 [26] Figs. 3a and 5 correspond to the case of narrow QWs.
The dependence of results from $\zeta=\nu/\nu_1$ is determined
only by there dependance from the coefficient $\nu_1$ at fixed
$\nu$ .}

\section{The reflectance, transmittance and absorbance at a pulse
irradiation.}

The energy flux ${\bf S}(p)$, where $p=t-z\nu_1/c$, corresponding
to the electric field of the stimulating pulse, is as follows
\begin{equation}\label{58}
    {\bf S}(p)={\bf e}_z{c\nu_1\over 4\pi}E_0^2(z, t)={\bf
    e}_zS_0P(p),~~~~~S_0={c\nu_1\over 2\pi}E_0^2.
\end{equation}
The transmitted flux on the right of the QW is
\begin{equation}\label{59}
    {\bf S}_{right}(p)={\bf e}_z{c\nu_1\over 4\pi}E_{right}^2(z, t)={\bf
    e}_zS_0\mathcal{T}(p).
\end{equation}
For the reflected flux (on the left of the QW) we obtain
\begin{equation}\label{60}
    {\bf S}_{left}(s)=-{\bf e}_z{c\nu_1\over 4\pi}(\Delta E_{left}(z, t))^2=-{\bf
    e}_zS_0\mathcal{R}(s).
\end{equation}
where $s=t+z\nu_1/c$. The dimensionless functions $\mathcal{T}(p)$
and $\mathcal{R}(s)$ determine the rates of the transmitted and
reflected energy of the stimulating pulse.

The time-dependance of the absorption is determined by the
dimensionless coefficient
\begin{equation}\label{61}
    \mathcal{A}(t)=P(t)-\mathcal{T}(t)-\mathcal{R}(t),
\end{equation}
if a time $t_{пр}$ of the pulse transition through the QW is small
in comparison to the pulse continuation $\Delta t$ (or if the
pulse length along the axes  $z$ is much more in comparison to the
QW's width $d$). Let us check that the criterium is fulfilled for
the wide enough QWs and pulses, duration of which is comparable
with the value  $\hbar/\Delta\omega$, where $\Delta\omega$ is the
distance between two polaron energy levels. Let us take, according
to [5], $\Delta\omega=0.0065 eV$ (the corresponding pulse duration
is $\Delta t=\hbar/\Delta\omega=0.1 ps$). The pulse time
transition through the QW with the width
$d=300$\AA~~~;$t_{пр}\simeq d\nu/с =10^{-4}\nu~~ps$ , i.e., the
condition $t_{пр}<<\Delta t$ is fulfilled, and the definition (60)
ia applicable.

Let us note that at a pulse irradiation the integral absorption
(integrated on time from $t=-\infty$ to $t=\infty$ ) equals 0, if
the non-radiative damping $\gamma=0$. For the monochromatic
irradiation the coefficients $\mathcal{T}, \mathcal{R}$ and
$\mathcal{A}$ are constants, and $\mathcal{A}=0$  at $\gamma=0$.

In [20,21,23-26] the light reflection and absorption are
considered for comparatively wide QWs ($kd\geq 1$). The results
(the formulas and graphics) were obtained with the help of the
"envelope" functions  $\phi_\chi(z)$ without taking into account
the Coulomb interaction of electrons and holes. In the present
work the calculations are performed without concretization the
"envelope" functions.The difference of the refraction coefficients
of a QW $\nu$  and barriers  $\nu_1$ is taken into account.
 \begin{figure}
 \includegraphics [] {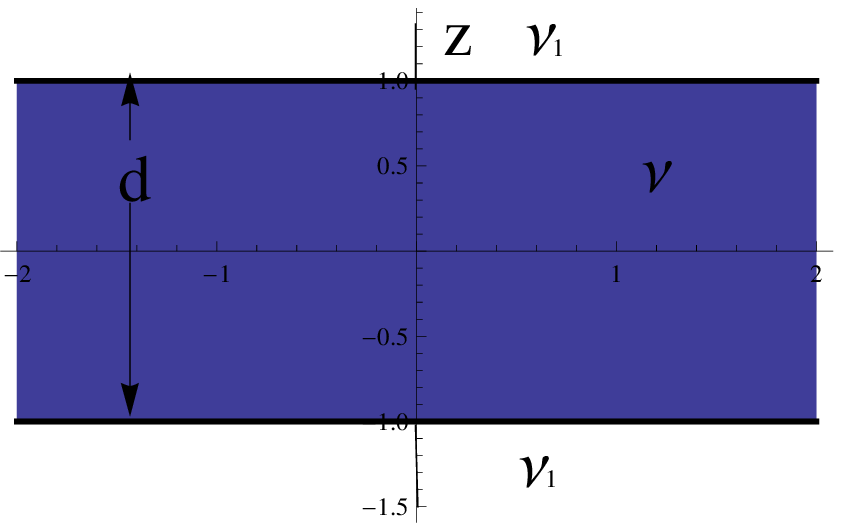} % Here is how to import EPS art
\caption[*]{\label{Fig.1.eps} }
 \end{figure}

\end{document}